\documentclass[aps,prd,twocolumn,nofootinbib]{revtex4-1}

\usepackage{graphicx}
\usepackage{hyperref}
\usepackage{slashed}
\usepackage{color}

\begin{document}

\title{Further Study on the Doubly Heavy Baryon Production around the $Z^0$ Peak at A High Luminosity $e^+ e^-$ Collider}

\author{Jun Jiang$^{1}$}
\author{Xing-Gang Wu$^{1}$}
\email{wuxg@cqu.edu.cn}
\author{Shao-Ming Wang$^{1}$}
\email{smwang@cqu.edu.cn}
\author{Jia-Wei Zhang$^{2}$}
\author{Zhen-Yun Fang$^{1}$}

\address{$^{1}$ Department of Physics, Chongqing University, Chongqing 401331, P.R. China \\
$^2$ Department of Physics, Chongqing University of Science and Technology, Chongqing 401331, P.R. China}

\date{\today}

\begin{abstract}

The doubly heavy baryon $\Xi_{QQ^{\prime}}$ ($Q^{(\prime)}$ = $b$ or $c$) is different from the ordinary baryons. The production of the doubly heavy baryon can provide valuable insight on how the colored $(QQ^{\prime})$-diquark can be transformed into the color-singlet baryon. We study the doubly heavy baryon production through the $e^+ e^-$ annihilation channel, $e^{+} + e^{-}\rightarrow\gamma/Z^0 \rightarrow \Xi_{QQ^{\prime}} +\bar{Q} +\bar{Q^{\prime}}$, within the nonrelativistic QCD framework. We calculate the baryon transverse momentum and the rapidity distributions for all these channels. Typical transverse momentum and rapidity cuts are adopted to show the properties of these distributions in detail. At a $e^+ e^-$ collider that runs around the $Z^0$-boson mass with a high luminosity up to ${\cal L} \simeq 10^{34-36}{\rm cm}^{-2} {\rm s}^{-1}$, in comparison to the Belle and BABAR experiments, it is found that sizable $\Xi_{cc}$, $\Xi_{bc}$ and $\Xi_{bb}$ events can be produced even after performing reasonable baryon transverse momentum and rapidity cuts. \\

\noindent {\bf PACS numbers:} 13.66.Bc, 12.38.Bx, 14.20.-c

\end{abstract}

\maketitle

Within the nonrelativistic QCD (NRQCD) framework~\cite{nrqcd}, the physical state of a heavy quarkonium is generally regarded as a superposition of Fock states, and the relative importance among those infinite ingredients is evaluated by the velocity scaling rule. The production of the doubly heavy baryon $\Xi_{QQ^{\prime}}$ can provide valuable insight on how the colored $(QQ^{\prime})$-diquark can be transformed into the color-singlet baryon. These baryons offer a good platform for testing various theories and models. Throughout the paper, if not specially stated, we shall ignore the isospin-breaking effect of the doubly heavy baryons; namely, the symbol $\Xi_{QQ'}$ is a short notation for all the baryon $\Xi_{QQ'q}$, where $Q^{(\prime)}$ = $b$ or $c$ quark and the light quark $q$ stands for $u$, $d$ or $s$, respectively.

In Ref.\cite{ee_baryon} we have calculated the doubly heavy baryon production through the $e^+ e^-$ annihilation, $e^{+} + e^{-}\rightarrow\gamma/Z^0 \rightarrow \Xi_{QQ^{\prime}} +\bar{Q} +\bar{Q^{\prime}}$, within the framework of NRQCD. We have discussed the total cross section for the doubly heavy baryon production through $e^+ e^-$ annihilation at the super-$Z$ factory~\cite{wjw}, a $e^+e^-$ collider with a high luminosity ${\cal L}\propto 10^{34-36}$cm$^{-2}$s$^{-1}$ and its colliding energy is around the $Z^{0}$ peak. Moreover, the International Linear Collider (ILC) is also programmed to run at the $Z^{0}$ peak with a luminosity of ${\cal L} = 0.7 \times 10^{34}$cm$^{-2}$s$^{-1}$ (GigaZ program~\cite{gigaz}). It has been observed that the $Z^0$-boson resonance effect shall raise the production rate by several orders of magnitude in comparison to the Belle and BABAR experiments, then it could be a better platform for studying the properties of heavy baryons. It is interesting to show more detailed production properties, such as how the baryon production cross section varies with its transverse momentum or rapidity, which is experimentally more useful.

According to NRQCD factorization theorem~\cite{nrqcd,petrelli}, the differential cross section of the $e^+ e^-$ annihilation process takes the following factorization form,
\begin{eqnarray}
&d\sigma(e^{+} e^{-}\rightarrow \Xi_{QQ^{\prime}}+\bar{Q}+\bar{Q^{\prime}}) \nonumber\\
&=\sum\limits_{n} d\hat\sigma \left(e^{+} e^{-}\rightarrow (QQ^{\prime})[n]+\bar{Q}+\bar{Q^{\prime}}\right) \langle{\cal O}^H(n)\rangle, \label{total}
\end{eqnarray}
where the matrix element $\langle{\cal O}^H(n)\rangle$ is proportional to the inclusive transition probability of the perturbative $(QQ^{\prime})[n]$ pair into the doubly heavy baryon $\Xi_{QQ^{\prime}}$, and the symbol $[n]$ represents the spin and color quantum numbers for the diquark state $(QQ^{\prime})$. Since the fragmentation function $D(z)$ for a heavy diquark into a baryon peaks near $z\approx 1$~\cite{K3,K2,K4}, the momentum of the final baryon may be considered roughly equal to the momentum of initial diquark. Thus, as an approximation, studying the hadronic production of $\Xi_{QQ^{\prime}}$ is equivalent to studying the hadronic production of $(QQ^{\prime})$-diquark. However, different to the heavy quark fragmentation, during the fragmentation of a diquark into a baryon, the diquark may dissociate \footnote{Due to the heavy quark-diquark symmetry \cite{qdi1,qdi2,qdi3}, there are relations between the doubly heavy baryons and the singly heavy mesons, e.g. they could have similar properties such as the spectrums and the productions or the decays. Then the doubly heavy diquarks may dissociate due to the radiation of soft partons with large probability as those of the heavy mesons.}, which will decrease the baryon production cross section to a certain degree. In the present paper, we will not take such dissociation effect into consideration. Regarding this point, our present estimations can be treated as an upper limit for the production channels. For the intermediate $(cc)$- and $(bb)$- diquarks, there are two spin and color configurations: $[^3S_1]_{\bf\bar{3}}$ and $[^1S_0]_{\bf 6}$; while for the $(bc)$-diquark, there are four spin and color configurations: $[^3S_1]_{\bf\bar{3}}$, $[^3S_1]_{\bf 6}$, $[^1S_0]_{\bf\bar{3}}$, and $[^1S_0]_{\bf 6}$.

The short-distance cross section
\begin{displaymath}
d\hat\sigma\left(e^{+} e^{-}\rightarrow (QQ^{\prime})[n]+\bar{Q}+\bar{Q^{\prime}}\right)  = \frac{\overline{\sum} |{\cal M}|^{2} d\Phi_3}{4\sqrt{(p_1\cdot p_2)^2-m_e^4}} , \label{cs}
\end{displaymath}
where ${\cal M}$ is the hard scattering amplitude, and $\overline{\sum}$ means we need to average over the spin states of the electron and positron and the sum over the color and spin of all final particles. The detailed treatment of ${\cal M}$ has been presented in Ref.\cite{ee_baryon}, the key point of which is to properly transform one of the fermion lines of the diquark production to an antifermion line of the meson production. $d{\Phi_3}$ stands for the conventional three-particle phase space, which can be generated and integrated totally or partly by using the FormCalc program~\cite{formcalc}, or a combination of RAMBOS~\cite{rambos} and VEGAS~\cite{vegas}, which can be found in the generators GENXICC~\cite{gencc} and BCVEGPY~\cite{bcvegpy}.

When doing the numerical calculation, we take the input parameters as~\cite{pdg}: $\Gamma_z=2.4952$ GeV, $m_Z=91.1876$ GeV, $m_W=80.385$ GeV and $m_H=125$ GeV. The renormalization scale for $\Xi_{QQ^{\prime}}$ as well as other input parameters, such as the bound state parameters and the non-perturbative matrix elements, are taken to be exactly the same as those adopted in Ref.~\cite{ee_baryon}.

\begin{table}[tb]
\caption{Total cross sections (in unit: pb) for the $\Xi_{QQ^{\prime}}(n)$ baryon production through $e^+ e^-$ annihilation at the $Z^{0}$ peak ($\sqrt{S}=m_Z$) or the Higgs peak ($\sqrt{S}=m_H$) respectively, where $n$ stands for the intermediate $(QQ^{\prime})$-diquark state. }
\begin{tabular}{|c||c||c|}
\hline
~~~ $$ ~~~ & ~$\sigma(\sqrt{S}=m_Z)$~ & ~$\sigma(\sqrt{S}=m_H)$~\\
\hline\hline
$e^+ e^-\to \gamma \to \Xi_{cc}([^3S_1]_{\bf\bar{3}})$  &  8.90 $\times$ $10^{-4}$  &  4.81 $\times$ $10^{-4}$\\
\hline
$e^+ e^-\to \gamma \to \Xi_{cc}([^1S_0]_{\bf 6}) $ &  4.29 $\times$ $10^{-4}$  &  2.32 $\times$ $10^{-4}$\\
\hline
$e^+ e^-\to Z^{0} \to \Xi_{cc}([^3S_1]_{\bf\bar{3}}) $  & 0.44  &  8.12 $\times$ $10^{-4}$\\
\hline
$e^+ e^-\to Z^{0} \to \Xi_{cc}([^1S_0]_{\bf 6}) $ &  0.21  &  3.94 $\times$ $10^{-4}$\\
\hline\hline
$e^+ e^-\to \gamma \to \Xi_{bc}([^3S_1]_{\bf\bar{3}}) $  & 2.66 $\times$ $10^{-4}$  &  1.45 $\times$ $10^{-4}$\\
\hline
$e^+ e^-\to \gamma \to \Xi_{bc}([^3S_1]_{\bf 6})$  & 1.33 $\times$ $10^{-4}$  &  7.24 $\times$ $10^{-5}$\\
\hline
$e^+ e^-\to \gamma \to \Xi_{bc}([^1S_0]_{\bf \bar{3}})$  &  1.85 $\times$ $10^{-4}$  &  1.05 $\times$ $10^{-4}$\\
\hline
$e^+ e^-\to \gamma \to \Xi_{bc}([^1S_0]_{\bf 6})$  &  9.27 $\times$ $10^{-5}$  &  5.27 $\times$ $10^{-5}$\\
\hline
$e^+ e^-\to Z^{0} \to \Xi_{bc}([^3S_1]_{\bf\bar{3}})$ & 0.61  &  1.15 $\times$ $10^{-3}$\\
\hline
$e^+ e^-\to Z^{0} \to \Xi_{bc}([^3S_1]_{\bf 6})$ & 0.30  &  5.73 $\times$ $10^{-4}$\\
\hline
$e^+ e^-\to Z^{0} \to \Xi_{bc}([^1S_0]_{\bf \bar{3}})$  & 0.44  &  8.43 $\times$ $10^{-4}$\\
\hline
$e^+ e^-\to Z^{0} \to \Xi_{bc}([^1S_0]_{\bf 6})$  & 0.22  &  4.21 $\times$ $10^{-4}$\\
\hline\hline
$e^+ e^-\to \gamma \to \Xi_{bb}([^3S_1]_{\bf\bar{3}})$  &  1.94 $\times$ $10^{-5}$  &  1.12 $\times$ $10^{-5}$\\
\hline
$e^+ e^-\to \gamma \to \Xi_{bb}([^1S_0]_{\bf 6})$  &  8.95 $\times$ $10^{-6}$  &  5.28 $\times$ $10^{-6}$\\
\hline
$e^+ e^-\to Z^{0} \to \Xi_{bb}([^3S_1]_{\bf\bar{3}})$  & 4.85 $\times$ $10^{-2}$  &  9.62 $\times$ $10^{-5}$\\
\hline
$e^+ e^-\to Z^{0} \to \Xi_{bb}([^1S_0]_{\bf 6})$  &  2.33 $\times$ $10^{-2}$  &  4.64 $\times$ $10^{-5}$\\
\hline
\end{tabular}
\label{crosssection}
\end{table}

\begin{figure}
\includegraphics[width=0.4\textwidth]{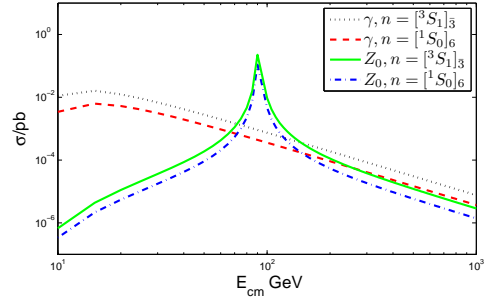}
\caption{Total cross section (in unit: pb) for the $\Xi_{cc}$-production channel, $e^{+}+e^{-}\rightarrow \gamma/Z^{0} \rightarrow \Xi_{cc}(n)+\bar{c}+\bar{c}$, versus the $e^+ e^-$ collision energy $E_{cm}=\sqrt{S}$, where $n$ stands for the intermediate $(cc)$-diquark state. Here the peak is at $\sqrt{S}=m_Z$. } \label{cc}
\end{figure}

Total cross sections for the $\Xi_{QQ^{\prime}}(n)$ baryon production through the $e^+ e^-$ annihilation at the $Z^{0}$ peak ($\sqrt{S}=m_Z$) or the Higgs peak ($\sqrt{S}=m_H$) are presented in Table \ref{crosssection} respectively, where $n$ stands for the intermediate $(QQ^{\prime})$-diquark state \footnote{We correct an input error for the parameter $g_w$ used in Ref.\cite{ee_baryon}, which should be $g_w=2\sqrt{2}m_W\sqrt{G_F/\sqrt{2}}$. Then, all the numerical results listed in Ref.\cite{ee_baryon} for the total cross-sections via the $Z^0$-propagator should be corrected and multiplied by an overall parameter $({m_W}/{m_Z})^4 = 0.6039$. All the Fortran sources are available upon request.}.

When the $e^+e^-$ collider runs at the $Z^{0}$ peak ($\sqrt{S}=m_Z$), the total cross section $\sigma(m_Z)$ for the production channel via the $\gamma$-propagator is about three orders lower than that of the channel via the $Z^0$-propagator. For the collider runs at the Higgs peak ($\sqrt{S}=m_H$), the total cross section $\sigma(m_H)$ via the $\gamma$-propagator is suppressed by only one order compared to that of the channel via the $Z^{0}$-propagator. As an explicit example, we present the total cross section for the production channel $e^{+}+e^{-}\rightarrow \gamma/Z^{0} \rightarrow \Xi_{cc}(n)+\bar{c}+\bar{c}$ versus the $e^+ e^-$ collision energy $E_{cm}=\sqrt{S}$ in Fig.(\ref{cc}). Moreover, we have
\begin{itemize}
\item By summing up all the intermediate $(QQ')$-diquark states' contributions into consideration, we obtain $\sigma_{tot}(m_Z)=0.65$ pb and $\sigma_{tot}(m_H)=1.92\times10^{-3}$ pb for $\Xi_{cc}$ production; $\sigma_{tot}(m_Z)=1.57$ pb and $\sigma_{tot}(m_H)=3.36\times10^{-3}$ pb for $\Xi_{bc}$ production; and $\sigma_{tot}(m_Z)=0.07$ pb and $\sigma_{tot}(m_H)=1.59\times10^{-4}$ pb for $\Xi_{bb}$ production.
\item In contrast to the BABAR and Belle, at the super-$Z$ factory, the production cross section for $\Xi_{bc}$ is about two times larger than that of $\Xi_{cc}$. Thus, the super $Z$ factory shall be helpful for studying the properties of $\Xi_{bc}$.
\item For the production channels through the $\gamma$-propagator, the cross sections at the $Z^0$ peak and the Higgs peak are on the same order of magnitude, i.e., about $10^{-6}$-$10^{-4}$ pb, which are small and can be safely neglected in comparison to the production channel through the $Z^{0}$-propagator.
\item As for the production through the $Z^{0}$-propagator, the cross sections $\sigma(m_H)$ are almost three orders smaller than $\sigma(m_Z)$ for all the mentioned doubly heavy baryons. This shows the super-$Z$ factory is much better than the super-Higgs factory for studying the doubly heavy baryon properties.
\end{itemize}

\begin{table}[htb]
\caption{The events of $\Xi_{QQ^{\prime}}(n)$ through the production channel $e^{+}e^{-} \rightarrow \gamma/Z^{0} \rightarrow \Xi_{QQ^{\prime}}(n)+\bar{Q}+\bar{Q^{\prime}}$ at various experiments, where $n$ stands for the intermediate diquark state. } \label{events}
\begin{tabular}{|c||c||c||c|}
\hline
  ~~~$$~~~    & ~~~super-$Z$~~~   & ~~~GigaZ~~~   & ~~super-Higgs~~ \\
\hline\hline
$\Xi_{cc}([^3S_1]_{\bf\bar{3}})$ & $4.40 \times 10^4$  & $3.08 \times 10^4$ & 129.3\\
\hline
$\Xi_{cc}([^1S_0]_{\bf 6})$ & $2.14 \times 10^4$  & $1.50 \times 10^4$  & 62.6\\
\hline\hline
$\Xi_{bc}([^3S_1]_{\bf\bar{3}})$ & $6.07 \times 10^4$  & $4.25 \times 10^4$ & 129.1\\
\hline
$\Xi_{bc}([^3S_1]_{\bf 6})$ & $3.03 \times 10^4$  & $2.12 \times 10^4$  & 64.5\\
\hline
$\Xi_{bc}([^1S_0]_{\bf \bar{3}})$ & $4.41 \times 10^4$  & $3.09 \times 10^4$  & 94.8\\
\hline
$\Xi_{bc}([^1S_0]_{\bf 6})$ & $2.21 \times 10^4$  & $1.55 \times 10^4$  & 47.4\\
\hline\hline
$\Xi_{bb}([^3S_1]_{\bf\bar{3}})$ & $4.86 \times 10^3$  & $3.40 \times 10^3$  & 10.7\\
\hline
$\Xi_{bb}([^1S_0]_{\bf 6})$ & $2.33 \times 10^3$  & $1.63 \times 10^3$  & 5.2\\
\hline
\end{tabular}
\label{tabmc}
\end{table}

Considering the super-$Z$ factory and the GigaZ program of ILC, we can estimate the $\Xi_{QQ^{\prime}}$ baryon events generated per year. Here, we adopt the luminosity of super-$Z$ factory as ${\cal L}\propto 10^{34}$cm$^{-2}$s$^{-1}$. Our estimations are put in Table~\ref{events}, where as a comparison, we also present the events at the super-Higgs factory \footnote{a $e^+e^-$ collider running at the Higgs peak with the same luminosity as that of the super-$Z$ factory, which has recently been suggested and stimulated by the discovery of the standard model Higgs-like particle at the LHC.}. Table~\ref{events} shows that both the super $Z$ factory and the GigaZ program can generate considerable baryon events, e.g. $\sim6.5\times10^4$ $\Xi_{cc}$ events, $\sim1.6\times10^5$ $\Xi_{bc}$ events and $\sim7.2\times10^{3}$ $\Xi_{bb}$ events can be generated at the super $Z$ factory; and the events per year under GigaZ are $0.7$ times of that of the super-$Z$ factory. Such a large number of events shall provide us great chances to study the properties of the doubly heavy baryons, such as to discuss its distributions and to further distinguish the baryons with different light quark contents, i.e., to distinguish $\Xi_{QQ^{\prime}u}$, $\Xi_{QQ^{\prime}d}$ and $\Omega_{QQ^{\prime}s}$.

\begin{figure}
\includegraphics[width=0.4\textwidth]{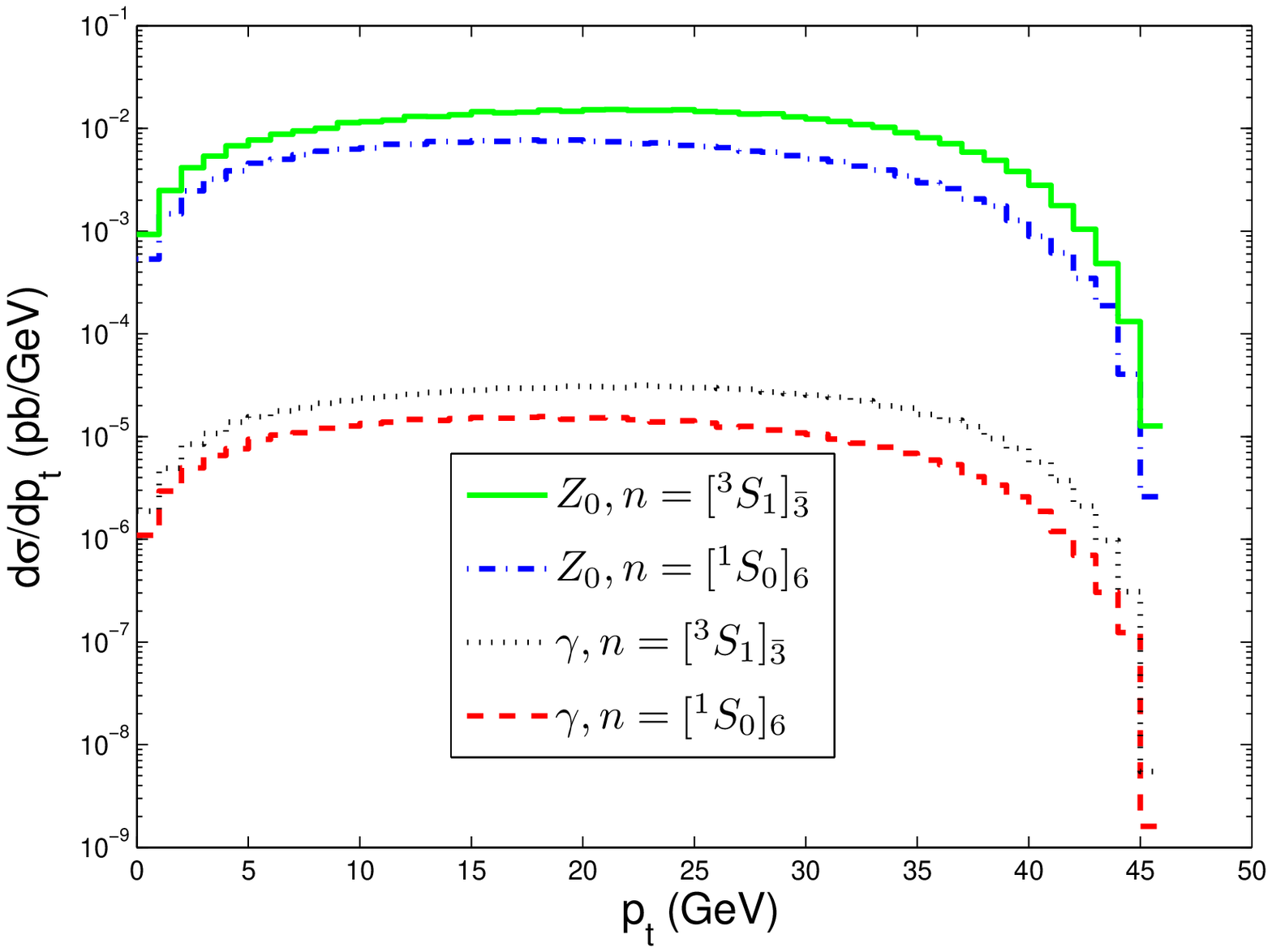}
\includegraphics[width=0.4\textwidth]{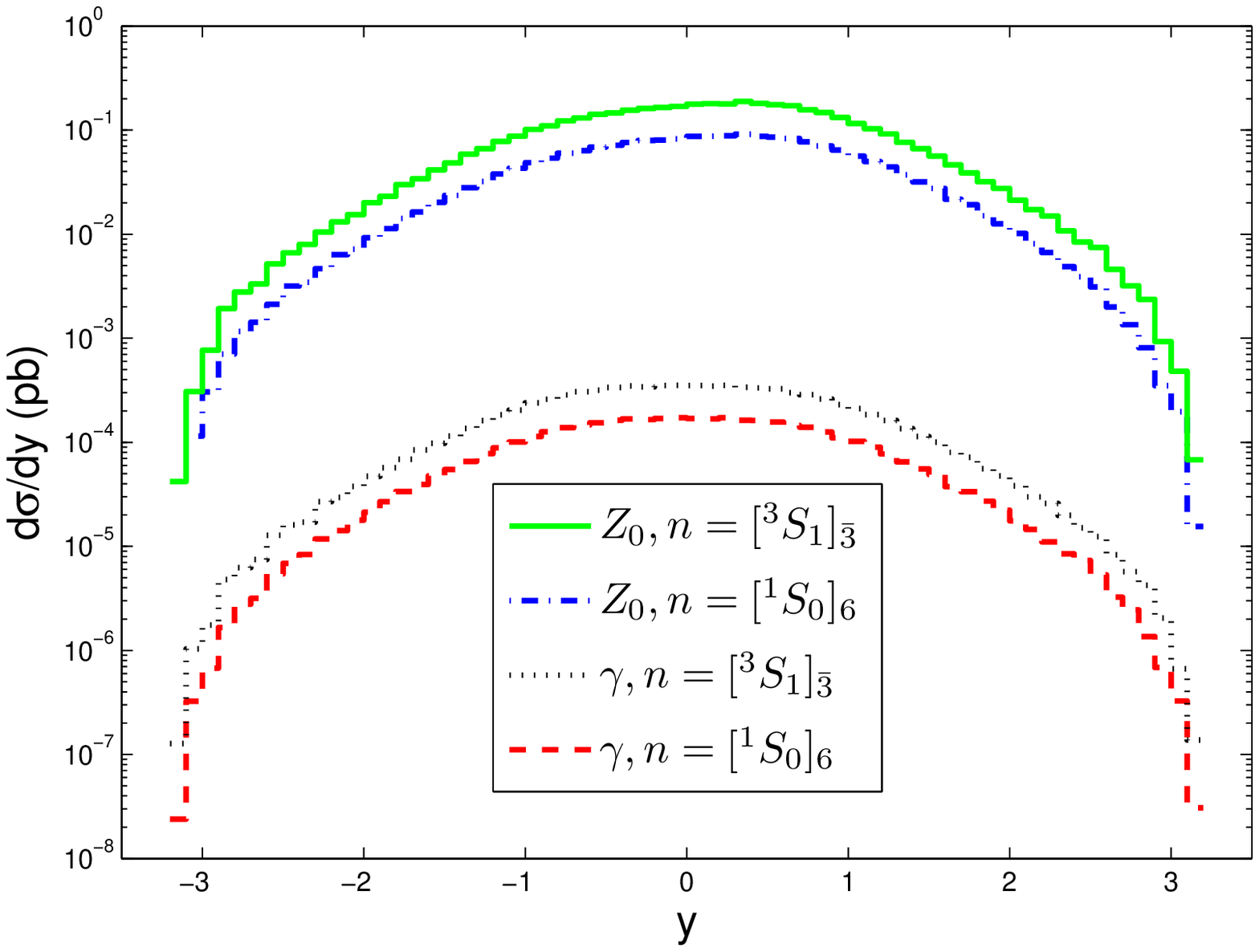}
\caption{The $p_t-$ and $y-$ distributions of the production channel $e^{+}+e^{-}\rightarrow \gamma/Z^{0} \rightarrow \Xi_{cc}(n)+\bar{c}+\bar{c}$ at $Z^0$ peak ($\sqrt{S}=m_Z$), where $n$ stands for the corresponding intermediate $(cc)$-diquark state. } \label{cc_pty}
\end{figure}

\begin{figure}
\includegraphics[width=0.4\textwidth]{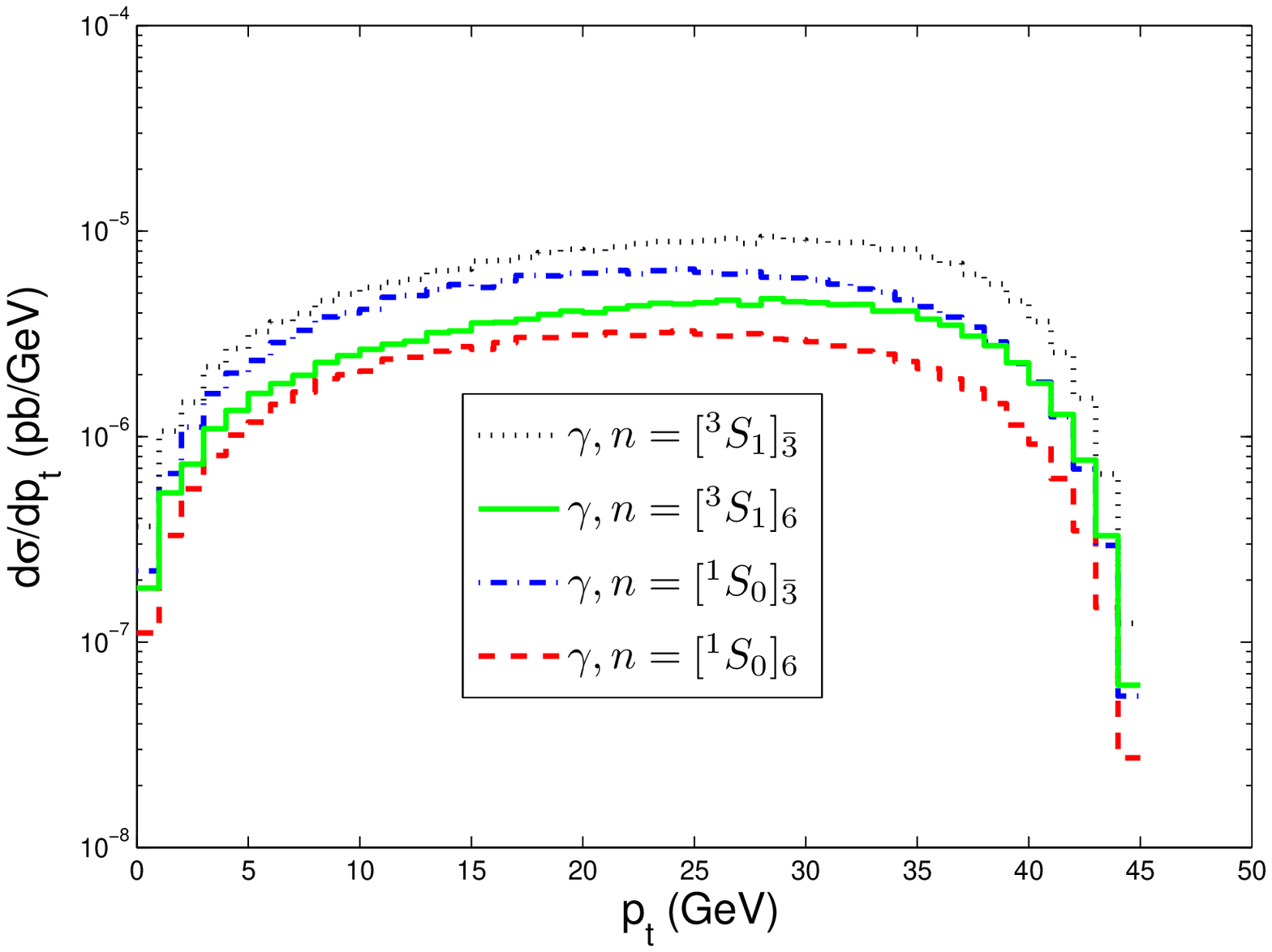}
\includegraphics[width=0.4\textwidth]{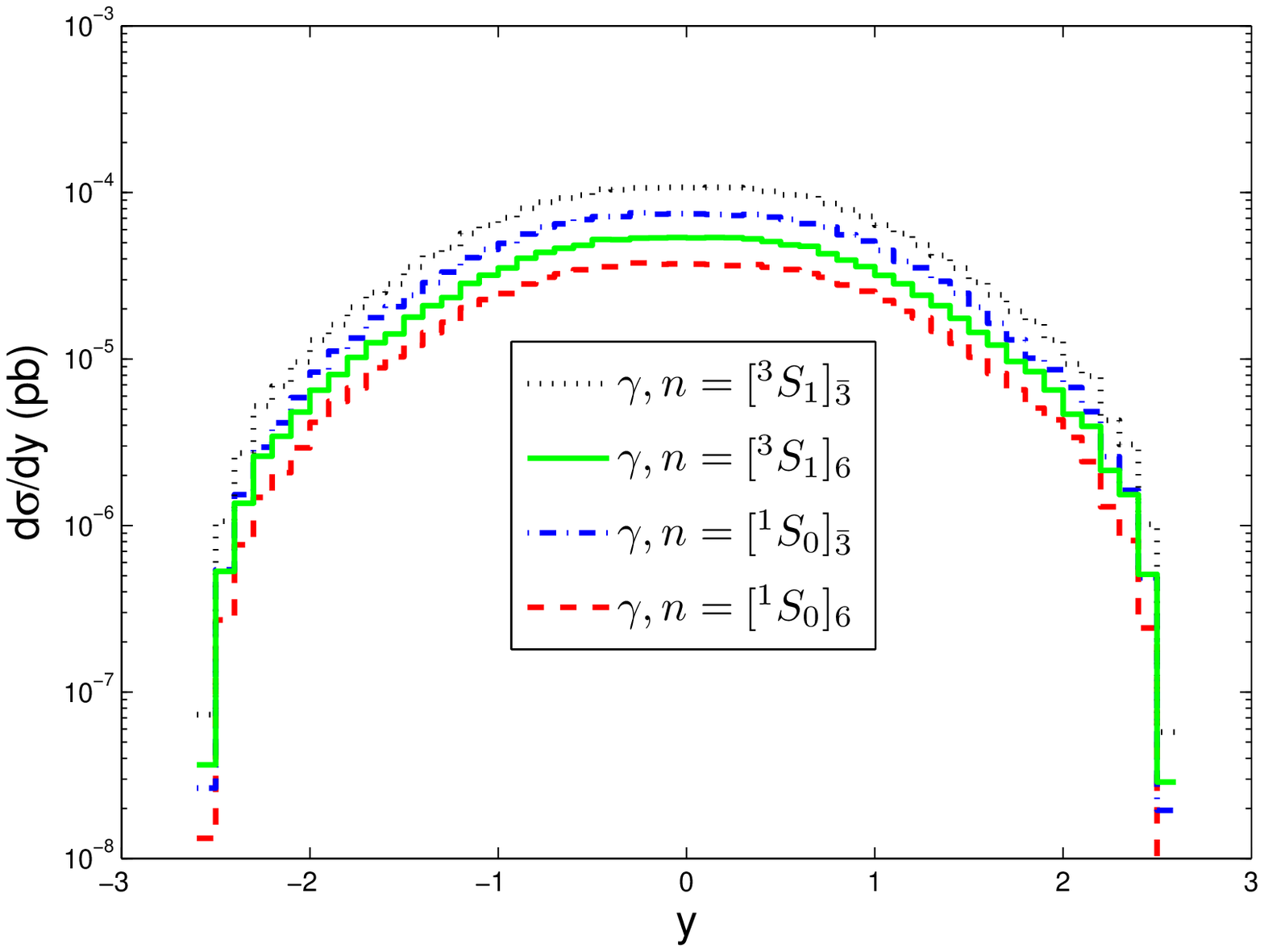}
\caption{The $p_t-$ and $y-$ distributions of the production channel $e^{+}+e^{-}\rightarrow \gamma \rightarrow \Xi_{bc}(n)+\bar{c}+\bar{b}$ at $Z^0$ peak ($\sqrt{S}=m_Z$), where $n$ stands for the corresponding intermediate $(bc)$-diquark state. } \label{bc_ptya}
\end{figure}

\begin{figure}
\includegraphics[width=0.4\textwidth]{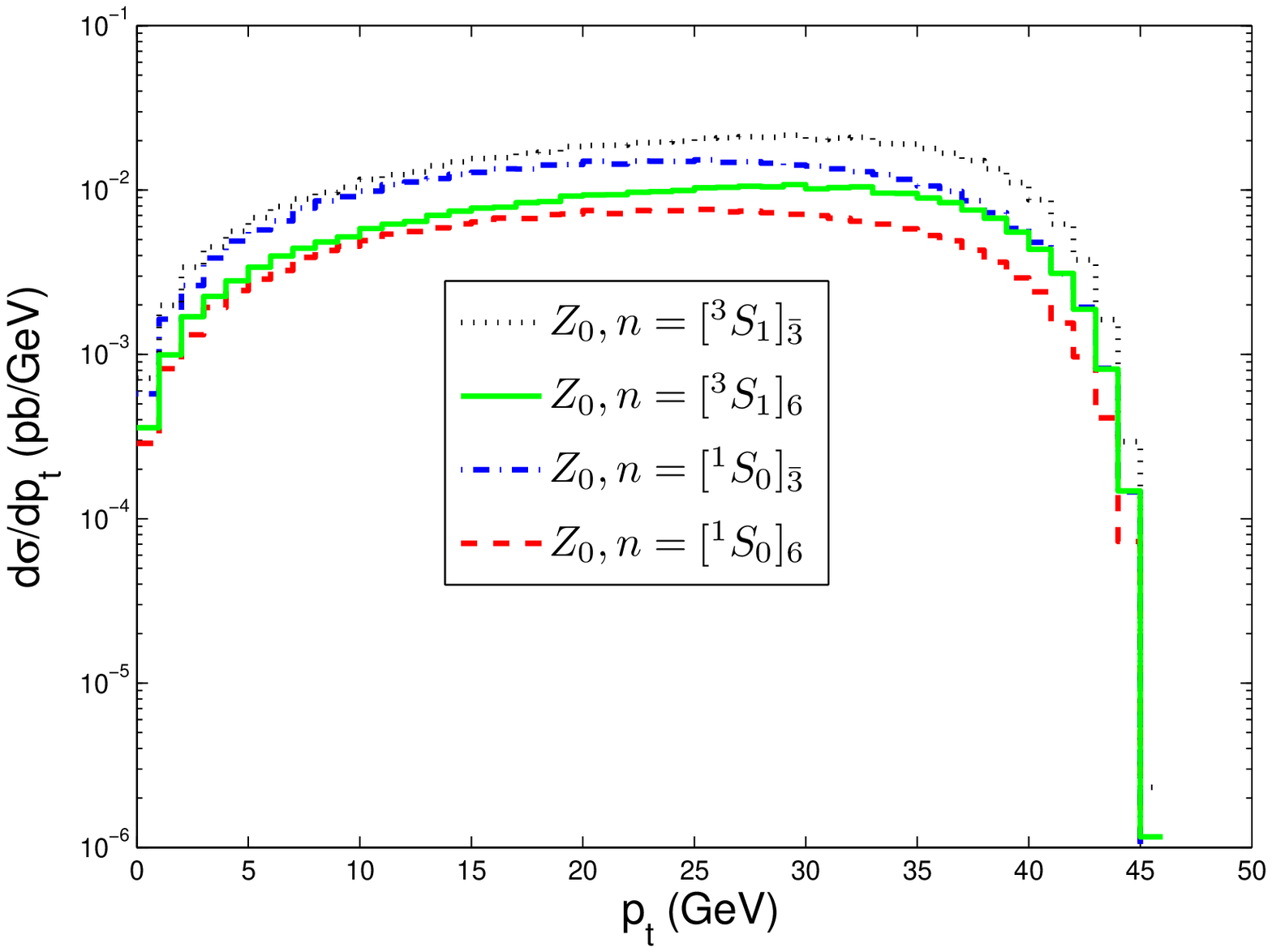}
\includegraphics[width=0.4\textwidth]{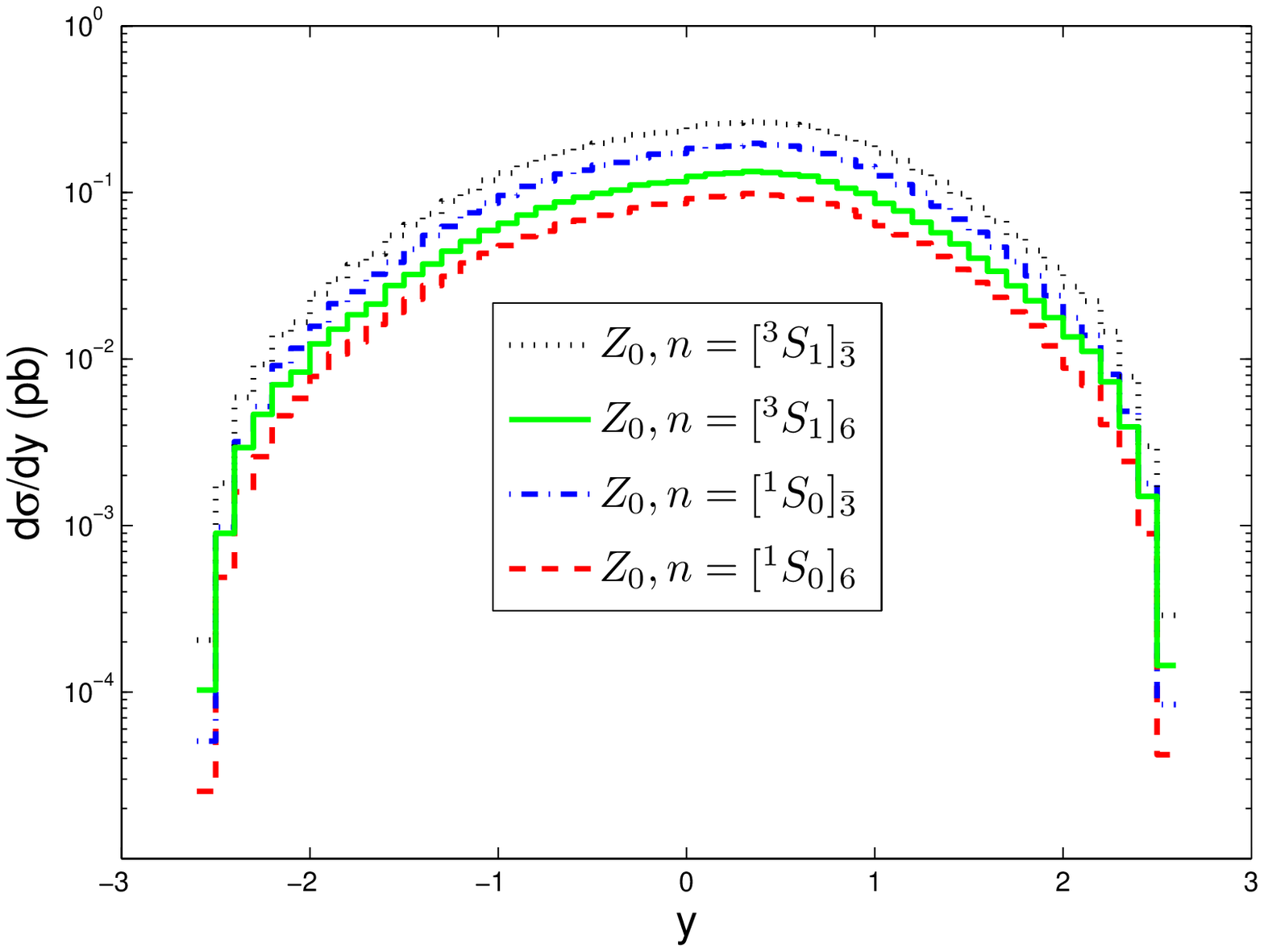}
\caption{The $p_t-$ and $y-$ distributions of the production channel $e^{+}+e^{-}\rightarrow Z^{0} \rightarrow \Xi_{bc}(n)+\bar{c}+\bar{b}$ at $Z^0$ peak ($\sqrt{S}=m_Z$), where $n$ stands for the corresponding intermediate $(bc)$-diquark state. } \label{bc_ptyb}
\end{figure}

\begin{figure}
\includegraphics[width=0.4\textwidth]{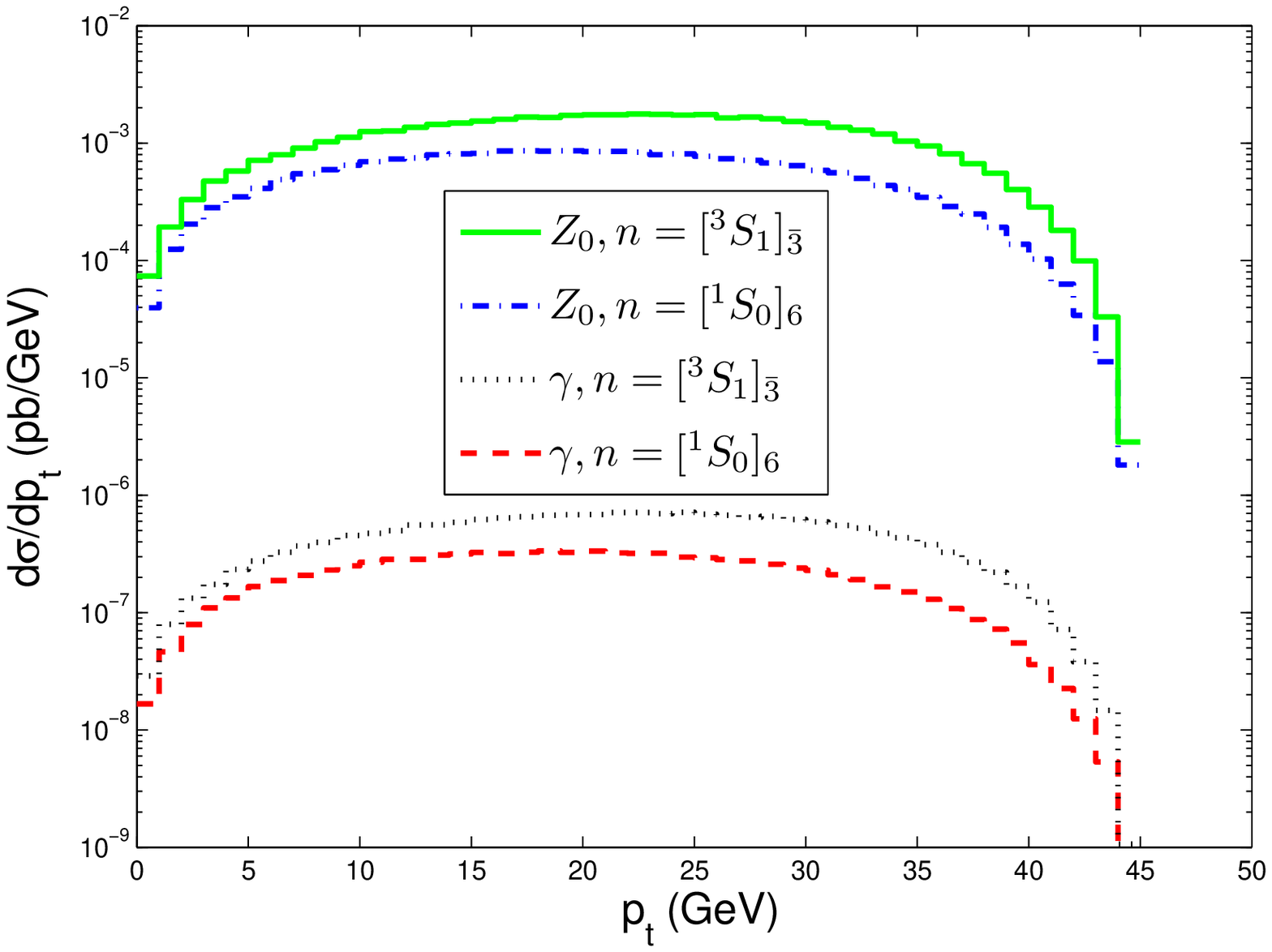}
\includegraphics[width=0.4\textwidth]{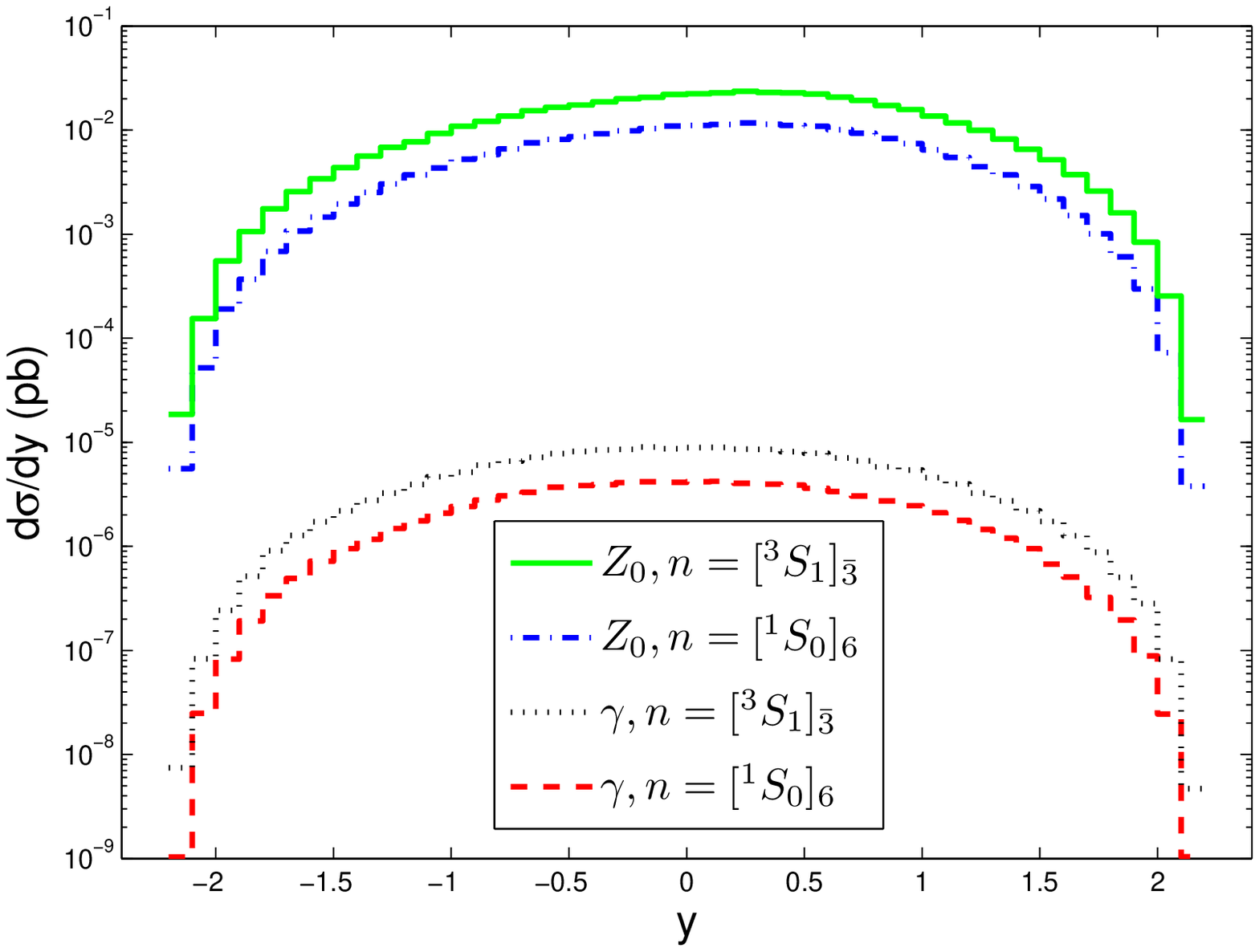}
\caption{The $p_t-$ and $y-$ distributions of the production channel $e^{+}+e^{-}\rightarrow \gamma/Z^{0} \rightarrow \Xi_{bb}(n)+\bar{b}+\bar{b}$ at $Z^0$ peak ($\sqrt{S}=m_Z$), where $n$ stands for the corresponding intermediate $(bb)$-diquark state. } \label{bb_pty}
\end{figure}

Next, we discuss the baryon transverse momentum ($p_t$) and rapidity ($y$) distributions for $\sqrt{S}=m_Z$. We present the $p_t$- and $y$- distributions for the production channels $e^{+}+e^{-}\rightarrow \gamma/Z^{0} \rightarrow \Xi_{QQ^{\prime}}(n)+\bar{Q}+\bar{Q^{\prime}}$ in Figs.(\ref{cc_pty},\ref{bc_ptya},\ref{bc_ptyb},\ref{bb_pty}), which are for $\Xi_{cc}(n)$, $\Xi_{bb}(n)$ and $\Xi_{bc}(n)$ respectively.

Experimentally, there is no detector that covers all the kinematics, only some of the events can be measured. Especially, the baryon events with a small $p_t$ and/or a large rapidity $y$, i.e. the produced baryons moving close to the beam direction, cannot be detected directly, so such kind of events cannot be utilized for experimental studies. Therefore, the baryon events with proper kinematic cuts on $p_t$ and $y$ must be put on precisely in the estimates. Considering detectors' abilities and to offer experimental references, we try various cuts in estimating the baryon production. We introduce the parameter $p_{tcut}$ with $p_{tcut}=A$ means $p_t \geq A$ and the parameter $y_{cut}$ with $y_{cut}=B$ means $|y| \leq B$.

\begin{table}[htb]
\caption{Total cross sections (in unit: pb) at the $Z^0$ peak ($\sqrt{S}=m_Z$) for the $\Xi_{QQ^{\prime}}$ baryon production through the process $e^{+}e^{-} \rightarrow\gamma/Z^{0}\rightarrow \Xi_{QQ^{\prime}}(n)+\bar{Q}+\bar{Q^{\prime}}$ versus the parameter $p_{tcut}$, where $n$ stands for the intermediate diquark state.} \label{ptcut}
\begin{tabular}{|c||c|c|c|c|c|}
\hline $p_{tcut}$ &~0 GeV~ & ~5 GeV~ & ~10 GeV~ & ~15 GeV~& ~20 GeV~ \\
\hline\hline $\sigma_{\Xi_{cc}([^3S_1]_{\bf\bar{3}})}$ & 0.440 & 0.420 & 0.373 & 0.310 & 0.236 \\
\hline $\sigma_{\Xi_{cc}([^1S_0]_{\bf 6})}$ & 0.214 & 0.201 & 0.174 & 0.139 & 0.101 \\
\hline total ($\sigma_{\Xi_{cc}}$) & 0.654 & 0.621 & 0.547 & 0.449 & 0.337 \\
\hline\hline$\sigma_{\Xi_{bc}([^3S_1]_{\bf\bar{3}})}$ & 0.607 & 0.590 & 0.546 & 0.480 & 0.397 \\
\hline $\sigma_{\Xi_{bc}([^3S_1]_{\bf 6})}$ & 0.303 & 0.295 & 0.273 & 0.240 & 0.198 \\
\hline $\sigma_{\Xi_{bc}([^1S_0]_{\bf \bar{3}})}$ & 0.441 & 0.427 & 0.390 & 0.333 & 0.265 \\
\hline $\sigma_{\Xi_{bc}([^1S_0]_{\bf 6})}$ & 0.221 & 0.214 & 0.195 & 0.167 & 0.133 \\
\hline total ($\sigma_{\Xi_{bc}}$) & 1.57 & 1.53 & 1.40 & 1.22 & 0.99 \\
\hline\hline $\sigma_{\Xi_{bb}([^3S_1]_{\bf\bar{3}})}$ & 0.0486 & 0.0468 & 0.0423 & 0.0355 & 0.0273 \\
\hline $\sigma_{\Xi_{bb}([^1S_0]_{\bf 6})}$ & 0.0233 & 0.0223 & 0.0196 & 0.0158 & 0.0116 \\
\hline total ($\sigma_{\Xi_{bb}}$) & 0.0719 & 0.0691 & 0.0619 & 0.0513 & 0.0389 \\
\hline
\end{tabular}
\end{table}

Total cross sections for the $\Xi_{QQ^{\prime}}$ baryon production at the $Z^0$ peak ($\sqrt{S}=m_Z$) under several typical transverse momentum cuts, $p_{tcut}=0$, $5$, $10$, $15$ and $20$ GeV, are put in Table \ref{ptcut}. With the increment of $p_{tcut}$, the total cross section $\sigma(\Xi_{cc})$ decreases by $5\%$, $12\%$, $18\%$ and $25\%$ in a step-by-step way; the total cross section $\sigma(\Xi_{bc})$ decreases by $3\%$, $8\%$, $13\%$ and $19\%$ respectively; and the total cross section $\sigma(\Xi_{bb})$ decreases by $4\%$, $10\%$, $17\%$ and $24\%$ respectively. It shows $\Xi_{cc}$ and $\Xi_{bb}$ have similar $p_t$ distributions, while $\Xi_{bc}$ is slightly different.

\begin{table}[htb]
\caption{Total cross sections (in unit: pb) at the $Z^0$ peak ($\sqrt{S}=m_Z$) for the $\Xi_{QQ^{\prime}}$ baryon production through the process $e^{+}e^{-} \rightarrow\gamma/Z^{0}\rightarrow \Xi_{QQ^{\prime}}(n)+\bar{Q}+\bar{Q^{\prime}}$ versus the parameter $y_{cut}$, where $n$ stands for the intermediate diquark state.} \label{ycut}
\begin{tabular}{|c||c|c|c|c|}
\hline $y_{cut}$ & ~~~2.0~~~ & ~~~1.5~~~ & ~~~1.0~~~ & ~~~0.5~~~ \\
\hline\hline $\sigma_{\Xi_{cc}([^3S_1]_{\bf\bar{3}})}$ & 0.424 & 0.389 & 0.310 & 0.171 \\
\hline $\sigma_{\Xi_{cc}([^1S_0]_{\bf 6})}$ & 0.206 & 0.189 & 0.151 & 0.083 \\
\hline total ($\sigma_{\Xi_{cc}}$) & 0.630 & 0.578 & 0.461 & 0.254  \\
\hline\hline$\sigma_{\Xi_{bc}([^3S_1]_{\bf\bar{3}})}$ & 0.593 & 0.546 & 0.434 & 0.239 \\
\hline $\sigma_{\Xi_{bc}([^3S_1]_{\bf 6})}$ & 0.297 & 0.273 & 0.217 & 0.119 \\
\hline $\sigma_{\Xi_{bc}([^1S_0]_{\bf \bar{3}})}$ & 0.433 & 0.400 & 0.319 & 0.175 \\
\hline $\sigma_{\Xi_{bc}([^1S_0]_{\bf 6})}$ & 0.217 & 0.200 & 0.159 & 0.088 \\
\hline total ($\sigma_{\Xi_{bc}}$) & 1.54 & 1.42 & 1.13 & 0.62 \\
\hline\hline $\sigma_{\Xi_{bb}([^3S_1]_{\bf\bar{3}})}$ & 0.0485 & 0.0461 & 0.0377 & 0.0214 \\
\hline $\sigma_{\Xi_{bb}([^1S_0]_{\bf 6})}$ & 0.0233 & 0.0224 & 0.0185 & 0.0106 \\
\hline total ($\sigma_{\Xi_{bb}}$) & 0.0718 & 0.0685 & 0.0562 & 0.0320 \\
\hline
\end{tabular}
\end{table}

Total cross sections for the $\Xi_{QQ^{\prime}}$ baryon production at the $Z^0$ peak ($\sqrt{S}=m_Z$) under several typical rapidity cuts, $y_{cut}=2.0$, $1.5$, $1.0$ and $0.5$, are put in Table \ref{ycut}. With the decrement of $y_{cut}$, total cross section $\sigma(\Xi_{cc})$ decreases by $8\%$, $20\%$ and $45\%$ respectively; total cross section $\sigma(\Xi_{bc})$ decreases by $8\%$, $20\%$ and $45\%$ respectively; total cross section $\sigma(\Xi_{bb})$ decreases by $5\%$, $18\%$ and $43\%$ respectively. It shows all the baryons have similar rapidity distributions.

As a summary,
\begin{itemize}
\item The SELEX collaboration~\cite{selex} has observed the $\Xi^+_{cc}$ events, however, neither BABAR nor Belle collaboration has found the evidence for $\Xi_{cc}$ in the related processes~\cite{Babar,Belle}. The CERN Large Hadronic Collider (LHC) or the future high luminosity $e^+e^-$ colliders (the super-$Z$ factory, the GigaZ project of ILC, and etc.) shall provide better platforms for measuring these baryons. Several studies on the heavy baryon production can be found in Refs.~\cite{FLSW,K1,K2,Michael,SPB,K3,Kiselev,cqww,chang,zhang,gencc,majp,ee_baryon,zhong,sizongguo}. Recently, it shows that even the triply heavy baryons could be measured at the LHC~\cite{triply}.

\item We have made a detailed study on the production properties of the doubly heavy baryon $\Xi_{QQ^{\prime}}$ through $e^+e^-$ annihilation. At the high luminosity super-$Z$ factory and the GigaZ project at the ILC, sizable events can be produced, which can provide a good platform for precise studies. More explicitly, at the super-$Z$ factory, we shall have $6.5\times10^4$ $\Xi_{cc}$, $1.6\times10^5$ $\Xi_{bc}$ and $7.2\times10^{3}$ $\Xi_{bb}$ events per year.

\item To be a useful reference, several typical $p_t$ and $y$ cuts have been adopted to show the baryon production properties in detail. These show that all the baryons have similar rapidity distributions, and the $\Xi_{cc}$ and $\Xi_{bb}$ baryons have similar $p_t$ distributions, while the $p_t$ distribution of $\Xi_{bc}$ is slightly different. Even applying a large $p_t$ cut up to $20$ GeV, we can still have sizable events that are about half of the case without any $p_t$ cut.

\item As a final remark, there are two dominant decay channels $\Xi^+_{cc} \to \Lambda^+_c ~\emph{K}^- ~\pi^+$ and $\Xi^+_{cc} \to \emph{p}~ \emph{D}^+ ~\emph{K}^-$ for $\Xi^+_{cc}$. Setting $\Gamma_{1,2}$ to be the decay widths of these two channels, we have~\cite{selex}: $$\Gamma_1/\Gamma_2 = 0.36 \pm0.21.$$ If taking $\Gamma_1/\Gamma_2 = 0.36$ and setting the relative possibility for various light quarks as $u :d :s \simeq 1:1:0.3$~\cite{pythia}, it is found that at the super-$Z$ factory with the luminosity ${\cal L}\propto 10^{34}$cm$^{-2}$s$^{-1}$, we can obtain 7.4 $\times$ $10^{3}$ $\Xi^+_{cc}$ events from the first channel and 2.1 $\times$ $10^{4}$ $\Xi^+_{cc}$ events from the second channel.
\end{itemize}

\hspace{2cm}

{\bf Acknowledgements}: This work was supported in part by the Fundamental Research Funds for the Central Universities under Grant No.CQDXWL-2012-Z002, by Natural Science Foundation of China under Grant No.11075225, No.11275280 and No.11205255, and by the Program for New Century Excellent Talents in University under Grant NO.NCET-10-0882.

\end{document}